\documentclass[12pt]{amsart}

\usepackage{amsmath}
\usepackage{amssymb}
\usepackage[all]{xy}           %For commutative diagrams
\usepackage{amssymb}           %For double arrows
\usepackage{fullpage}
\usepackage{hyperref}
\usepackage{eucal}
\usepackage{graphicx}
\usepackage{float}
\usepackage{epsfig}
\usepackage[usenames,dvipsnames]{color}
\usepackage{charter}
\usepackage{color}
\usepackage{comment}
\usepackage{hyperref}
\usepackage{fancyvrb} %For Verbatim boxes

\usepackage{subcaption}

\begin{document}
\title{Socialbots supporting human rights}
\author{E. Vel\'azquez, M. Yazdani, P. Su\'arez-Serrato}
%\\
%IMUNAM, CALIT2 UCSD, IMUNAM
%}

\begin{abstract}
Socialbots, or non-human/algorithmic social media users, have recently been documented as competing for information dissemination and disruption on online social networks. Here we investigate the influence of socialbots in Mexican Twitter in regards to the ``Tanhuato" human rights abuse report. We analyze the applicability of the BotOrNot API to generalize from English to Spanish tweets and propose adaptations for Spanish-speaking bot detection. We then use text and sentiment analysis to compare the differences between bot and human tweets. Our analysis shows that bots actually aided in information proliferation among human users. This suggests that taxonomies classifying bots should include non-adversarial roles as well.  Our study contributes to the understanding of different behaviors and intentions of automated accounts observed in empirical online social network data. Since this type of analysis is seldom performed in languages different from English, the proposed techniques we employ here are also useful for other non-English corpora.
%\keywords{social network analysis, socialbots, human rights, Spanish, Mexico}
\end{abstract}
\maketitle
%\noindent

As of 2017, Twitter has over 318 million monthly ``active users" \cite{twitter_users} - an amount that is more than population of Indonesia, the 4th most populous country in the world. Advances in Artificial Intelligence, however, has made it possible to automate the creation of online social media accounts that attempt to behave similarly to human users. These non-human accounts are known as {\it socialbots} or simply bots. A range of intentions and goals drive the production and deployment of such bots. Often, socialbots intervene in the discussion of specific trending topics to potentially manipulate, deceive, and  distract human users (for one review, see \cite{RiseSocialBots2016}). 

While there have been numerous studies on the impact and influence of socialbots, most previous studies have been limited to English Twitter. In this paper we present a case study to see how social bots are used in Mexican Twitter on a specific trending topic. We followed over 20 social events in Mexican Twitter in 2016, covering topics ranging from political scandals, attacks against the media, journalists, expressions of homophobia, to the banal and trivial. We found that the topic related to the report documenting the violation of human rights in Tanhuato had far more bot activity than other topics. We therefore focus our scope of study to tweets related to the \#Tanhuato hashtag that was trending in relation to the release of this report. We now give a brief background of this hashtag and topic. 

\subsection{Background}
As part of the war on drugs on May 22\textsuperscript{nd}, 2015 the Mexican armed forces raided a ranch in Tanhuato, Michoac\'an. After an extensive investigation, the National Commission for Human Rights (Comisi\'on Nacional de Derechos Humanos CNDH) released a report on August 18th, 2016. They established that at least 22 civilians were arbitrarily executed, victims suffered instances of torture, and that the crime scene was tampered with.

The report from the CNDH was made available online, and they used their Twitter account to promote access to it \footnote{Report available at \\ {\tt http://www.cndh.org.mx/sites/all/doc/Recomendaciones/ViolacionesGraves/RecVG\_004.pdf}}. During the following days there was an increased interest in the topic in Mexican Twitter, using the hashtag \#Tanhuato. We collected over 20K tweets using Twitter's streaming API between the 19th and 21st of August 2016. These tweets were processed and the user ID's evaluated with {\it BotOrNot} \cite{BoN2016} immediately after collection\footnote{Predecessor of {\it Botometer}.}. 

We found a substantive presence of socialbots using the \#Tanhuato hashtag during the collection period. According to {\it BotOrNot}\cite{BoN2016}, out of a total of 9,730 unique accounts we found high bot scores for 1,777 accounts. By following the retweets of the total collection of users we found an additional 26 bot accounts, giving us a total of 1,803 bots detected. Given this significant bot activity, we investigate what is the intention behind such bot accounts and their impact on spreading or stifling information. Since most text and bot analysis is typically done with English corpora, we also adapt our analysis for tweets in Spanish.

Unexpectedly, we found from our analysis that in fact most of the bot accounts were not acting maliciously and were in fact promoting access to the CNDH report. Human users retweeted the content of the tweets generated by bots, so that access to this report proliferated through the support of the socialbots and in coordination with the human users that retweeted them. What was the intention of the bots that we detected using \#Tanhuato ? We shall argue that they were helping to provide access to the report issued by the CNDH. This type of behavior sets them apart from the typically observed bots that have spam, or even censorship, intentions \cite{Wooley2016}, \cite{KPM2013}. %Such phenomena of potential censorship via flooding of a hash-tag has been previously observed in Mexican twitter \cite{suarez2016influence}. 

It is important to pause here and notice that in an instance like this it is not a clear matter whether these bots were benevolent or malignant. It is a matter of perspective. From the point of view of the Mexican armed forces, these bots are acting against their honor. From the point of view of the CNDH they are promoting access to a report of human rights abuse. Our study thus provides an interesting empirical test case for social bots acting as promoters, as opposed to suppressors, of information.

\subsection{Previous work}

Correlations of content between different accounts has also been used as a twitter bot detection technique \cite{chavoshi2016identifying}. Tweet sentiment has been studied to discriminate human from non-human accounts \cite{Dickerson2014}. Other methods combine graph-theoretic, syntactic, and semantic features to find bots \cite{Chu2012}. Another method to identify bots exploits natural language processing  \cite{Clark2015}. The possibility of creating a call to arms for activists using Twitter has been previously explored, and in fact this case study seems to be a variation on this theme \cite{Savage2015}. Numerous other previous works have addressed the issues of detection and classification of bots, see for example \cite{Wang2010,Dickerson2014,Hu:2013:SSD:2540128.2540508,Thomas:2011:SAR:2068816.2068840,Yang:2011:USN:2068816.2068841,Zhu:2012:DSS:2900728.2900753,Lee11sevenmonths,Ratkiewicz_truthy:mapping,Thomas:2013:TFA:2534766.2534784,Lee:2014:EFE:2948303.2948599,Beutel:2013:CSG:2488388.2488400,Hu:2014:SSD:2760438.2760604,Boshmaf:2013:DAS:2450570.2450801}.

%Bots promoting vaporizers and e-cigarettes and their effect on the public have been recently researched \cite{Clark2016}.

Most of the mentioned methods and previous results have been developed for English. By using the language-independent features of \textit{BotOrNot} \cite{BoN2016} it is possible to flag potential bot accounts in Spanish, and in other languages as well.

\section{Bot identification and data preparation}

\begin{figure}[ht]
\includegraphics[width=\textwidth]{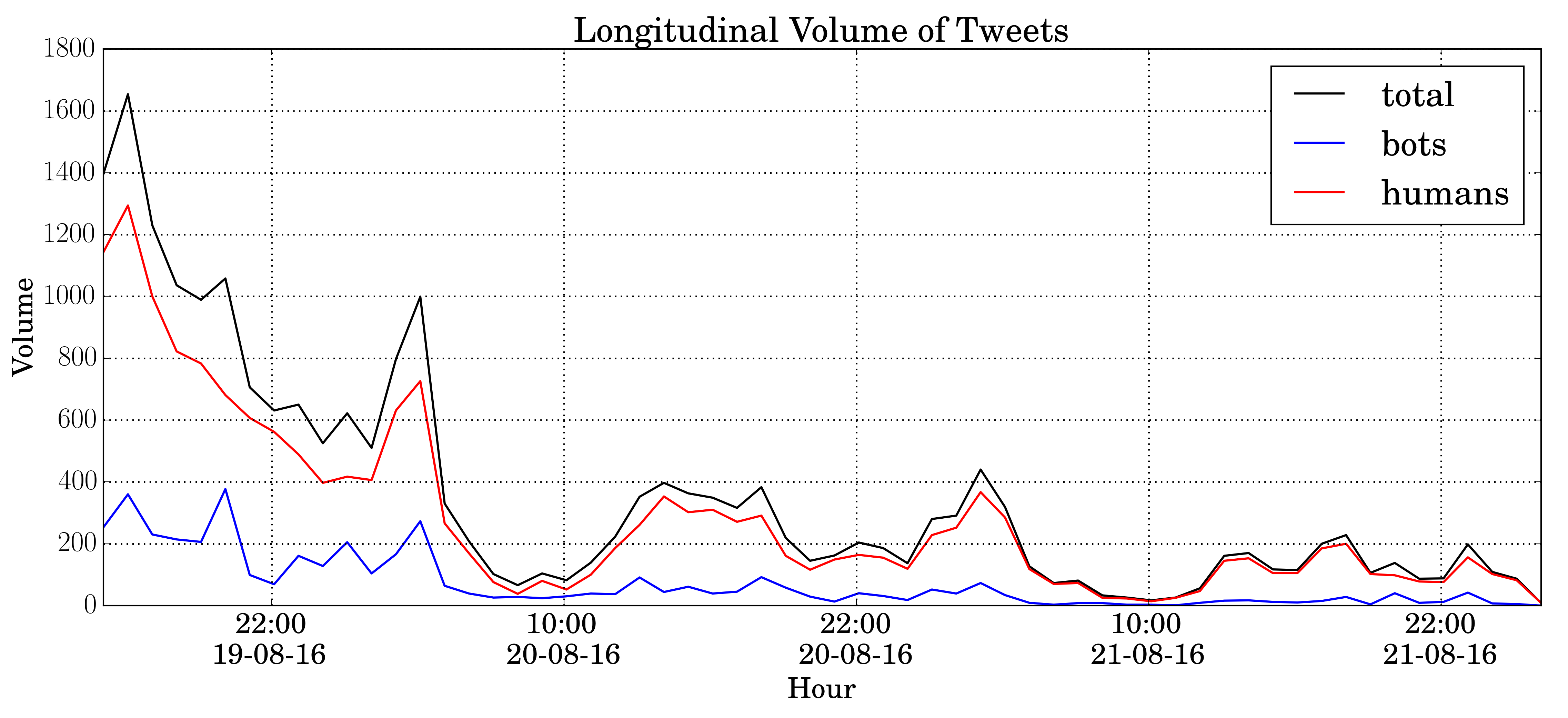}
\caption{ Bot versus human activity using \#Tanhuato, from streamed tweets during collection period.\label{fig:longitudinal_volume_tweets}} 
\end{figure}

In order to detect bots we use {\it BotOrNot}, a general supervised learning system designed for detecting socialbot accounts on Twitter \cite{BoN2016}. It utilizes over 1,000 features such as user meta-data, social contacts, diffusion networks, content, sentiment, and temporal signatures. Based on evaluation on a large set of labeled accounts, {\it BotOrNot} is extremely accurate in distinguishing bots from humans accounts, with an Area Under the ROC Curve (AUC) of 94\%.

When a twitter account is evaluated in {\it BotOrNot}, the output is a JSON file with several scores. As we are examining a corpus of tweets in Spanish we focus on language-independent classifiers, which show a large number of potential bot accounts. Surprisingly, combining the results of these language-independent classifiers is sufficient for detecting bots in Spanish. This suggests that simply discarding the language-dependent features of {\it BotOrNot} can yield to non-English bot detection. Further research should be done to validate the transferability of {\it BotOrNot} outside of English Twitter.

\begin{figure}
\centering
\includegraphics[width=.8\textwidth]{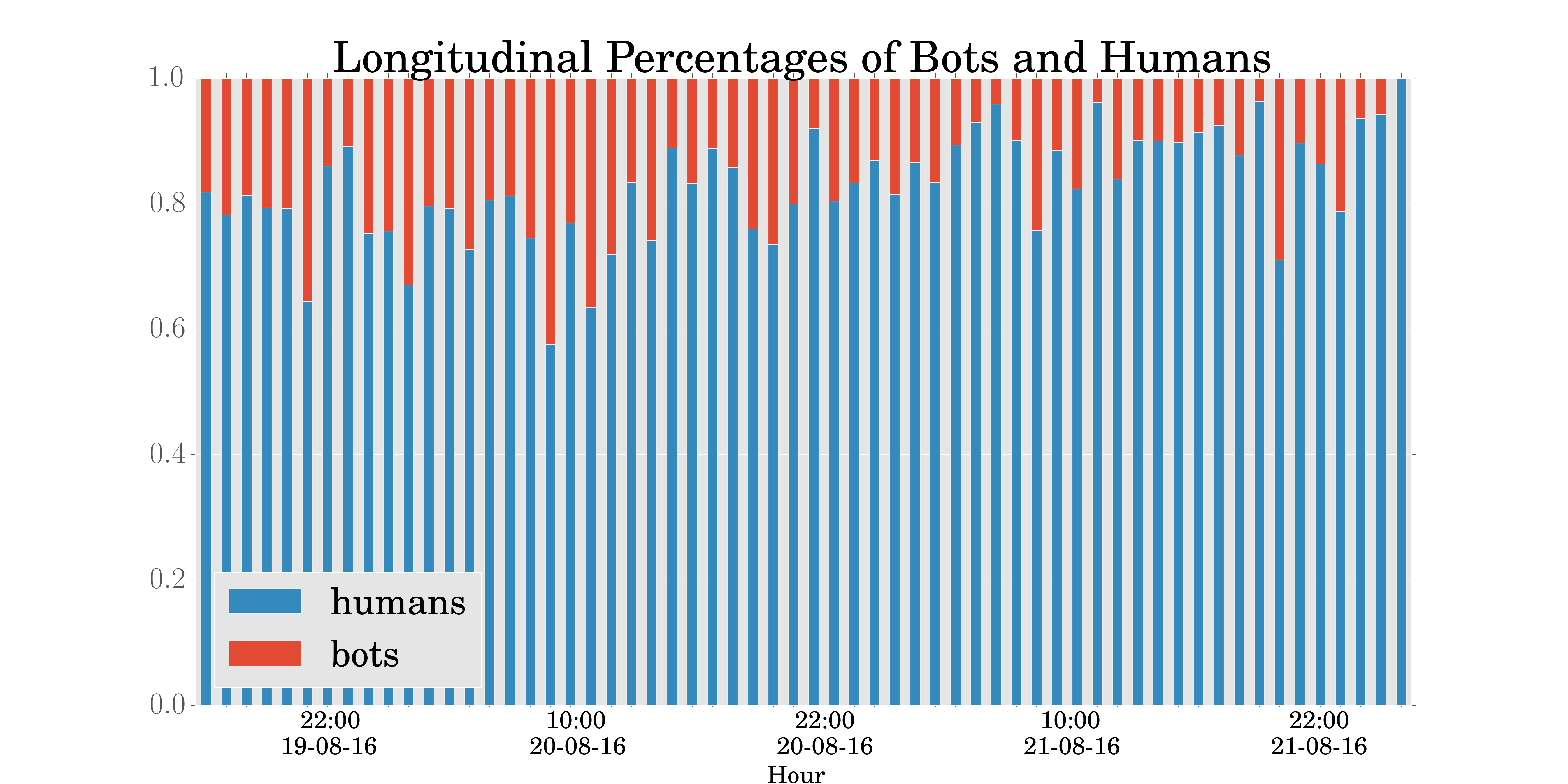}
\caption{ Percentages of Bot versus human activity using \#Tanhuato, from streamed tweets during the collection period.\label{fig:percent-bot-human}} 
\end{figure}

\begin{figure}[ht]
\centering
\includegraphics[width=.8\textwidth]{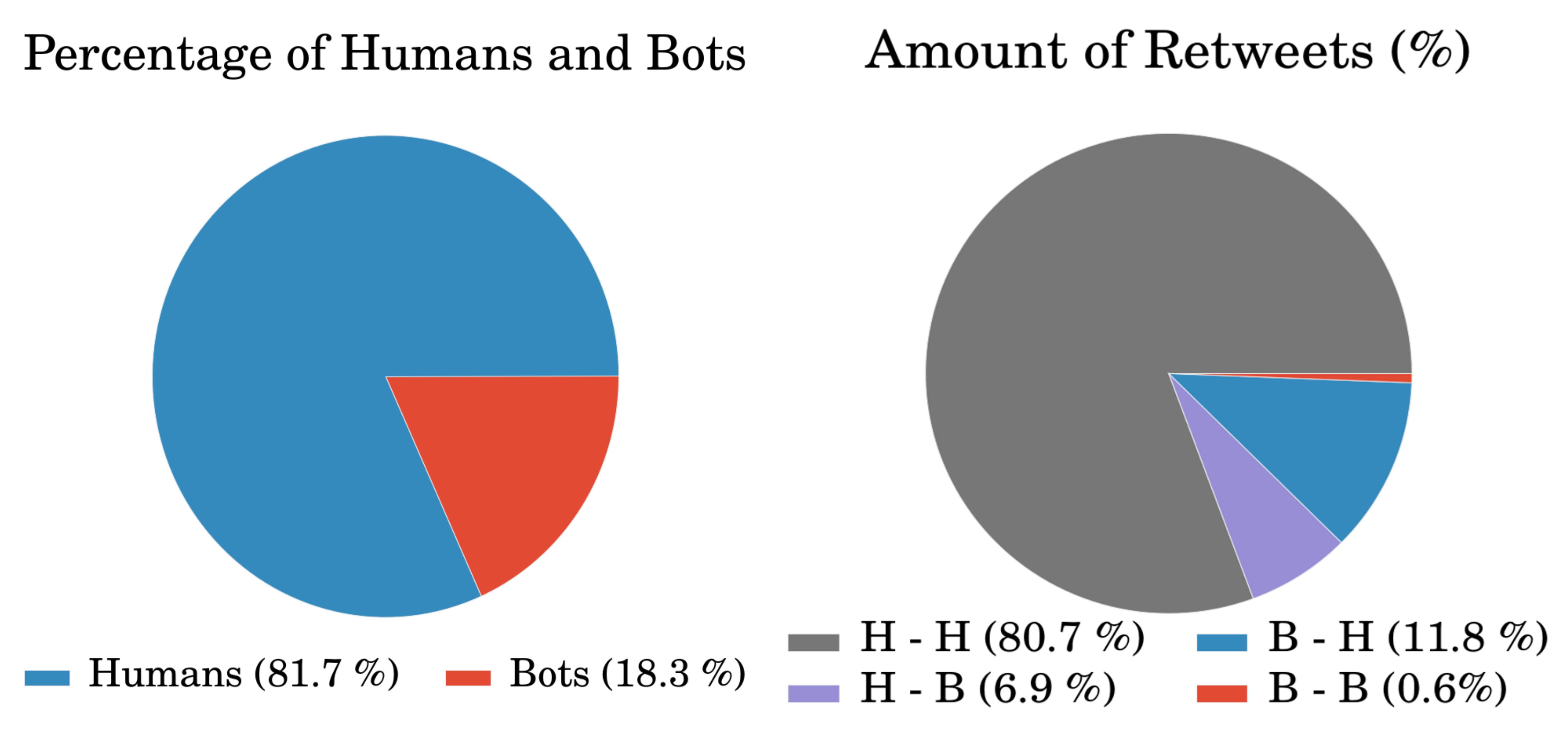}
\caption{{\sc Left:} Percentage of different human and bot accounts in collected data. Volume of registered retweets by user type. {\sc Right: }Clasification is as follows: humans retweeting humans ({\bfseries H-H}), bots retweeting humans.\label{fig:pie1}}
\end{figure}

We streamed 20,854 tweets from Twitter's API between 2016-08-19 15:06:17 and 2016-08-22 02:13:35. These tweets were generated by 9730 different users (see Figures \ref{fig:longitudinal_volume_tweets} and \ref{fig:pie1} for the relation between humans and bots), and among them we have 12905 retweets. When a user (human or bot) generates a tweet, and this tweet can be retweeted by a bot or a human. Consequently, we find four possibilities: a tweet created by a human and retweeted by another human ({\bfseries H-H}), created by a human and retweeted by a bot ({\bfseries H-B}), created by bot and retweeted by human ({\bfseries B-H}) or bot ({\bfseries B-B}). In Figure \ref{fig:longitudinal_volume_tweets} we show the evolution of \#Tanhuato in the collection period. The percentages of accounts that are humans and those that are bots are shown in Figure \ref{fig:percent-bot-human}. 
%; among these users we found 1777 bots and 7953 humans. 

\begin{figure}[ht]
\centering
\includegraphics[width=\textwidth]{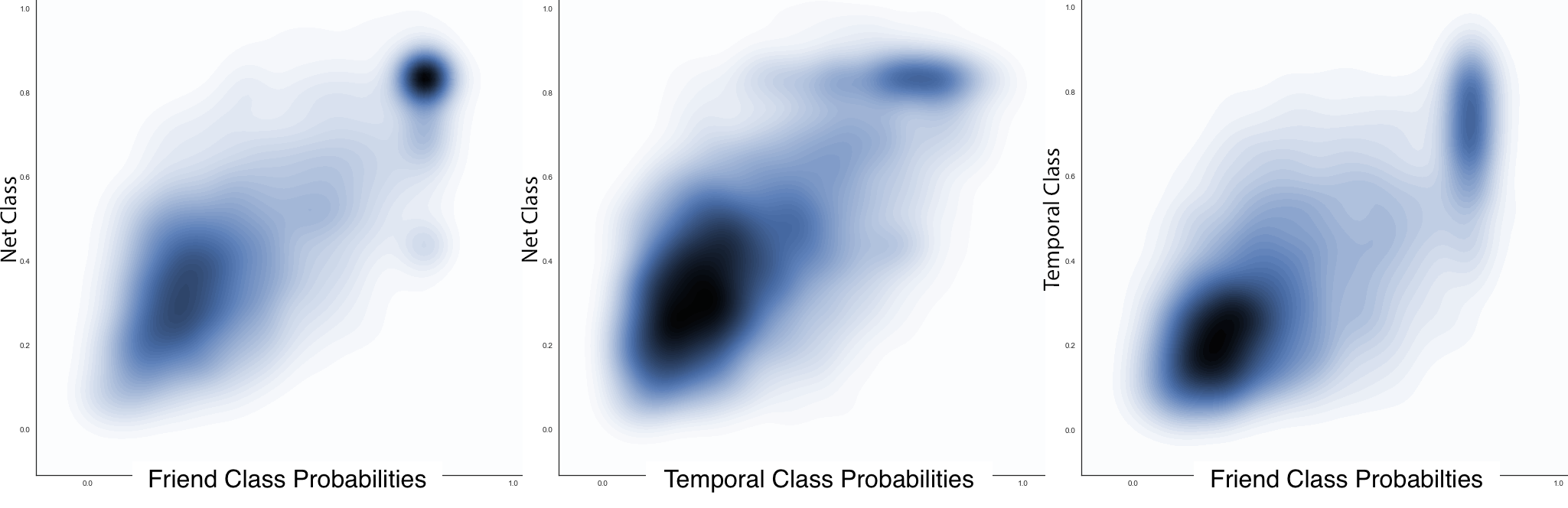}
\caption{ 2D Kernel decomposition estimate for Friend, Network, and Temporal classifiers from Bot-Or-Not, for \#Tanhuato, 19-21st August 2016, sample obtained through Twitter's streaming API.\label{fig:kde_subfigs}} 
\end{figure}

\begin{figure}[ht]
\centering
%\begin{minipage}{.45\textwidth}
\begin{center}
\centering
\includegraphics[width=0.6\textwidth]{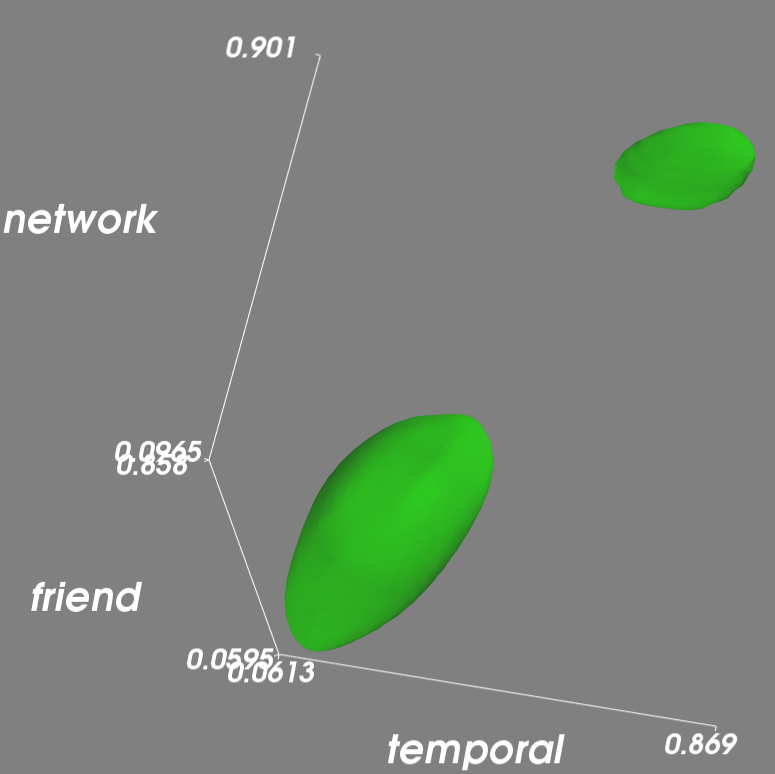}
\caption{ 3D Kernel decomposition estimate for Friend, Network, and Temporal classifiers from Bot-Or-Not, for \#Tanhuato, 19-21st August 2016, sample obtained through Twitter's streaming API.\label{fig:3d-kde-friend-net-temporal}} 
\end{center}
%\end{figure}
% \begin{center}
% \centering
% \includegraphics[width=\textwidth]{bar.png}
% \caption{ Percentages of Bot versus human activity using \#Tanhuato, from streamed tweets during the collection period.\label{fig:percent-bot-human}} 
% \end{center}

%\end{minipage}\hfill
% \begin{minipage}{.4\textwidth}
%   \centering
% \includegraphics[width=.4\textwidth]{pie1_rescaled.png}
% \caption{{\sc Up:} Percentage of different human and bot accounts in collected data. Volume of registered retweets by user type. {\sc Down: }Clasification is as follows: humans retweeting humans ({\bfseries H-H}), bots retweeting humans.\label{fig:pie1}}
%   \end{minipage}
  \end{figure}

In Figure \ref{fig:kde_subfigs} we show the bi-variate kernel decomposition estimates for pairwise combinations of the Friend, Network, and Temporal classifiers form {\it BotOrNot}. The regions towards the upper right hand corner correspond to areas where the bot scores are high. It can be clearly seen how the bot accounts naturally cluster. The final visualization of this analysis is presented in Figure \ref{fig:3d-kde-friend-net-temporal}, where we now compute the kernel density estimate that incorporates the three classifiers Friend, Network, and Temporal. In this image the smaller cluster in the upper right corner is the region where the bots accumulate. This 3D image is formed by taking iso-surfaces obtained from the 3D kernel density estimate. Again, as in the 2D images, we can separate the bot accounts in a natural way, to isolate them for further analysis. Notice that these three classifiers are all non-language specific and this is the reason behind focusing on them instead of on the overall bot score produced by {\it BotOrNot}. Having identified the bots present in our sample, we can now understand how the appeared over the collection period, as shown in Figure \ref{fig:percent-bot-human}.

% NEXT FIGURE NOT IN AIES

\begin{figure}
\centering
\includegraphics[width=\textwidth]{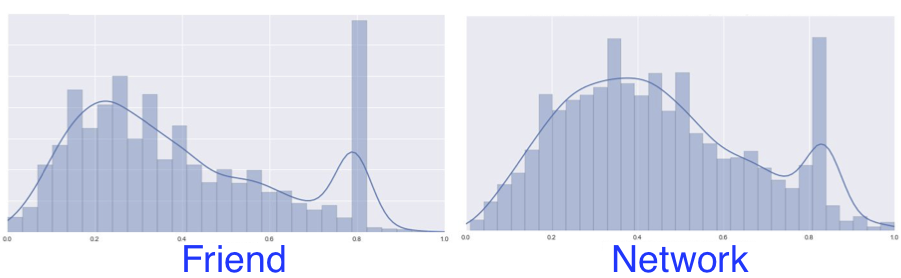}
\caption{ Kernel decomposition estimate for Friend (left) and Network (right) from Bot-Or-Not, for \#Tanhuato, 19-21st August 2016, sample obtained through Twitter's streaming API.\label{friend-n-net-dist}} 
\end{figure}

% \begin{figure}[ht]
% \label{fig:2d-kde-friend-net}
% \centering
% \includegraphics[width=\textwidth]{KDE-Friend-Net-Tanhuato_19-20_8_16.png}
% \caption{ 2D Kernel decomposition estimate for Friend and Network classifiers from Bot-Or-Not, for \#Tanhuato, 19-21st August 2016, sample obtained through Twitter's streaming API.} 
% \end{figure}

% \begin{figure}[ht]
% \label{fig:2d-kde-friend-temporal}
% \centering
% \includegraphics[width=\textwidth]{KDE-Friend-Temporal-Tanhuato_19-20_8_16.png}
% \caption{ 2D Kernel decomposition estimate for Friend and Temporal classifiers from Bot-Or-Not, for \#Tanhuato, 19-21st August 2016, sample obtained through Twitter's streaming API.} 
% \end{figure}

% \begin{figure}[ht]
% \label{fig:2d-kde-net-temporal}
% \centering
% \includegraphics[width=\textwidth]{KDE-Net-Temporal-Tanhuato_19-20_8_16.png}
% \caption{ 2D Kernel decomposition estimate for Friend and Network classifiers from Bot-Or-Not, for \#Tanhuato, 19-21st August 2016, sample obtained through Twitter's streaming API.} 
% \end{figure}

 % NEXT FIGURE NOT IN AIES
 
% NEXT FIGURE NOT IN AIES

\section{Network Analysis}
%%%%% % FROM HERE NOT IN AIES
\begin{center}
\begin{table}[htb]
\setlength{\tabcolsep}{8pt}
\centering
\begin{tabular}{lll}
54649261	& 0.0011778193	& H  \\ 
2903265492	& 0.0006699197	& H  \\
3060823412	& 0.0005209566	& H  \\
249005175	& 0.0004625333	& H  \\
1190644922	& 0.0004146598	& H  \\
163552910	& 0.0002160787	& H  \\
222959337	& 0.0001811033	& H  \\
78941875	& 0.000122997	& H  \\
\end{tabular}
\caption{Table of accounts with highest betweenness-centrality from the full retweet network. Notice that all of these accounts are from human users. The third column is labeled H for human and B for bot.}
\end{table}
\end{center}
%%%%

\begin{center}
\begin{table}[htb]
\setlength{\tabcolsep}{8pt}
\begin{center}
\begin{tabular}{lll}
3243658266  &   	787   &   	B\\
54649261   &   	754   &   	H\\
163552910   &   	594   &   	H\\
84613584   &   	471   &   	B\\
520653311  &   	438   &   	H\\
435299501   &   	368   &   	H\\
35977487   &   	328   &   	H\\
318799346   &   	212   &   	H\\
252160277   &   	211   &   	H\\
1911952410   &   	196   &   	H\\
\end{tabular}\hspace{1.7cm} \begin{tabular}{lll}
44554692   &   	191   &   	H\\
132346487   &   	156   &   	H\\
244218738   &   	154   &   	H\\
832309426182901760   &   	141   &   	H\\
825966216   &   	140   &   	H\\
200932969   &   	131   &   	H\\
18430394   &   	121   &   	H\\
296592711   &   	119   &   	H\\
43115590   &   	119   &   	H\\
190143362   &   	114   &   	H\\
 & &
\end{tabular}
\caption{Table of accounts with highest betweenness-centrality from the full retweet network. Notice that only two are marked as bots, @pictoline and @Pajaropolitico, both of these accounts belong to news organizations. The third column is labeled H for human and B for bot.}
\end{center}
\end{table}
\end{center}

\begin{table}[htb]
\setlength{\tabcolsep}{8pt}
\begin{center}
\begin{tabular}{l l l}
3243658266   &   	768   &   	B   \\
84613584   &   	444   &   	B   \\
52998787   &   	81   &   	B   \\
33884545   &   	71   &   	B   \\
91430932   &   	27   &   	B   \\
358862898   &   	26   &   	B   \\
357050985   &   	20   &   	B   \\
28608099   &   	16   &   	B   \\
104683173   &   	15   &   	B   \\
22721695   &   	11   &   	B   \\
2605229921   &   	9   &   	B   \\
558251048   &   	9   &   	H   \\
\end{tabular}\hspace{1.7cm} \begin{tabular}{lll}
93797343   &   	9   &   	B   \\
3122019163   &   	8   &   	B   \\
343452977   &   	8   &   	B   \\
319883780   &   	7   &   	B   \\
3907628182   &   	7   &   	B   \\
2372256601   &   	6   &   	B   \\
266390655   &   	5   &   	B   \\
755873792250023936   &   	4   &   	B   \\
3406088807   &   	4   &   	H   \\
85123108   &   	4   &   	B   \\
4251942192   &   	4   &   	H   \\
\end{tabular}
\caption{Table showing highest degree nodes in the retweet network. The third column is labeled H for human and B for bot.}
\end{center}
\end{table}

% % TO HERE NOT IN AIES

\begin{figure}
\centering
\begin{subfigure}{.5\textwidth}
  \centering
  \includegraphics[width=.5\linewidth]{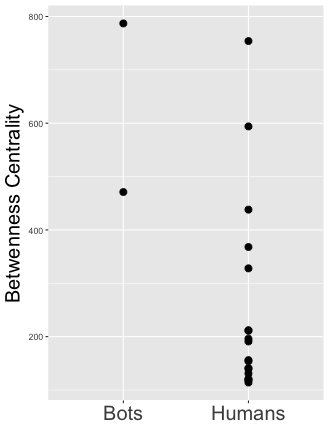}
  \caption{Retweet network betweenness centrality: the two bots are news organizations.}
  \label{fig:sub1}
\end{subfigure}%
\begin{subfigure}{.5\textwidth}
  \centering
  \includegraphics[width=.5\linewidth]{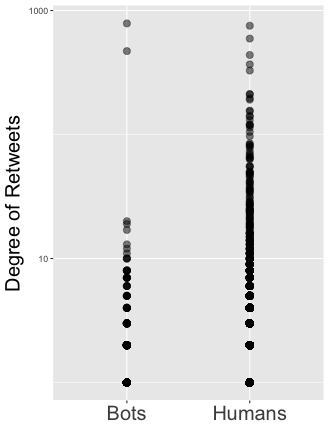}
  \caption{Highest degree nodes (raw retweet counts) in the retweet network.}
  \label{fig:sub2}
\end{subfigure}
\caption{Distribution of centrality of bot and human Twitter accounts. . We only show the top Twitter accounts.}
\label{fig:test}
\end{figure}

Now that we have performed our bot analysis, we can analyze the bot and human Twitter network. In Figure \ref{fig:sub1} we see that the nodes with the highest betweenness centrality in the full retweet network are all human, except for two accounts that belong to bots. These bot accounts are in fact official news organizations @pictoline and @Pajaropolitico. Thus, by the betweenness centrality in the retweet network, human users constitute the shortest paths of dialogue. With the exception of the formal news bots, socialbots are not playing an active role in the retweet network. 

Figure \ref{fig:sub2} shows the number of retweets by each user (measure of degree in the retweet network) and that again humans are the more active retweeters. In Figure \ref{fig:pie1} ({\sc down}) we find the relation between these quantities for our data. Furthormore, we observed in the data that the bots with the highest ammount of retweets among humans were mainly news organizations: @pictoline, @Pajaropolitico, @emeequis, @CNNEE, and @NewsweekEspanol. 

% FROM HERE NOT IN AIES

\begin{figure}
\centering
\includegraphics[scale=0.80]{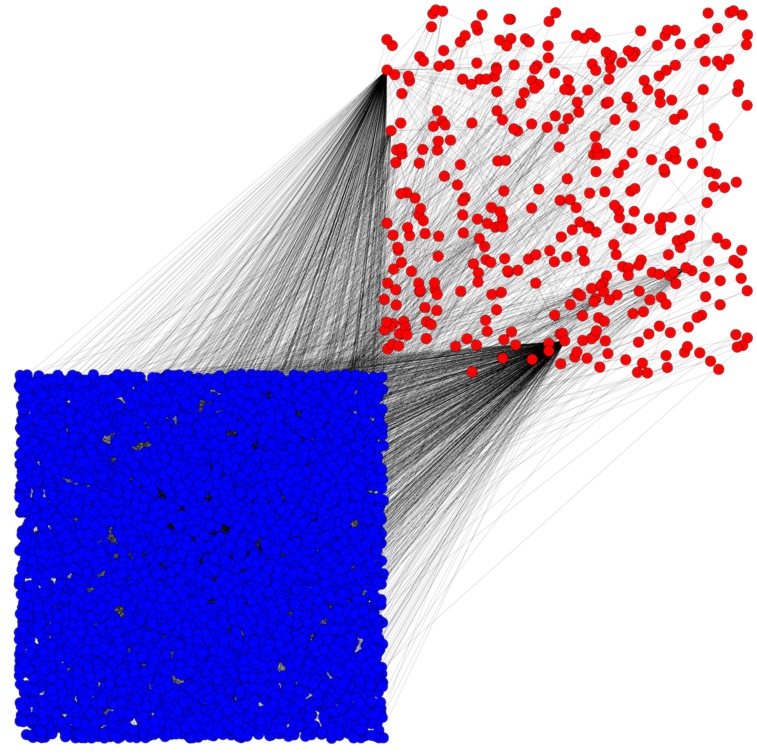}
\caption{Retweet network for \#Tanhuato, bots in red, humans in blue. Total of 6,528 nodes, and 10,011 edges.\label{fig:RT_net}}
\end{figure}

\begin{figure}
\centering
\includegraphics[width=\textwidth]{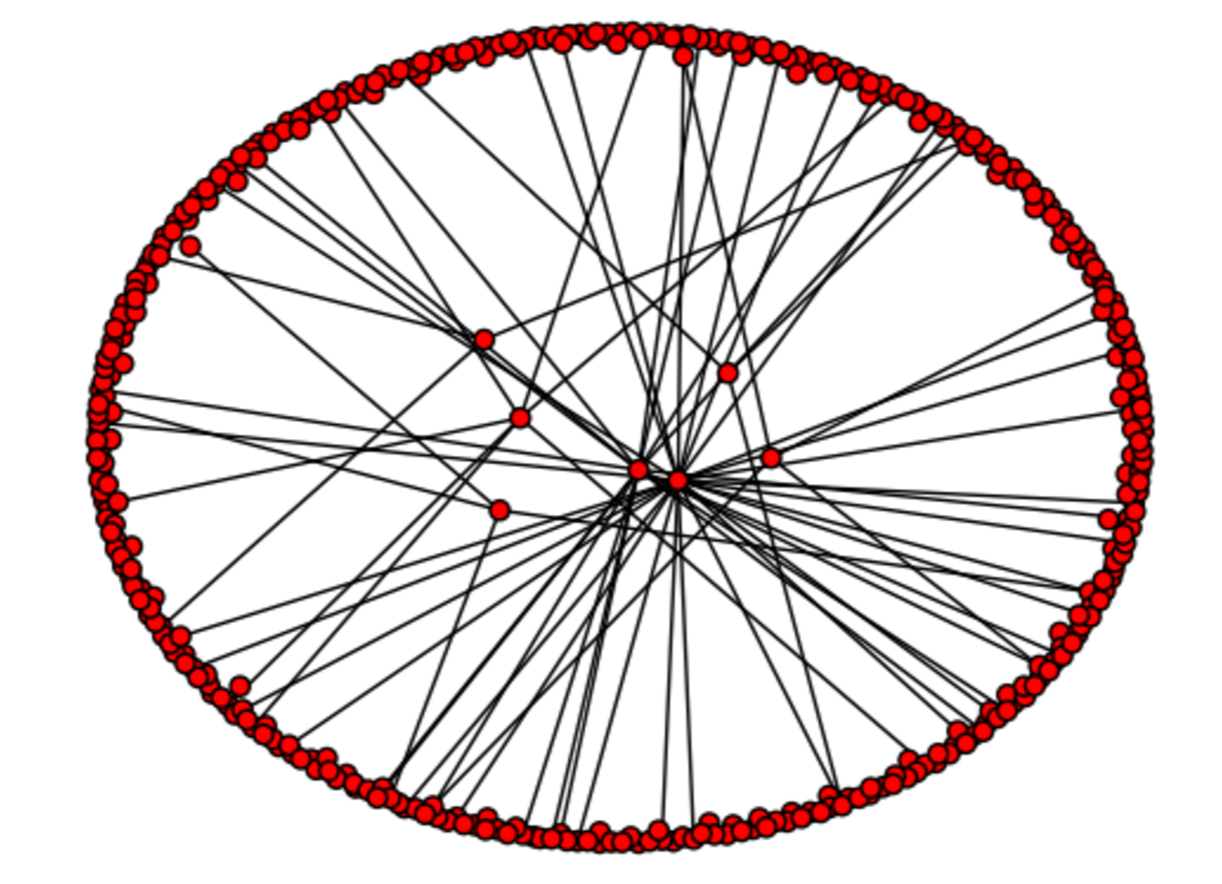}
\caption{Retweet network for \#Tanhuato obtained from our sample, bots retweeting bots. All edges are shown, most of the nodes in the outer circle have no connecting edge. This network is composed of 92 nodes, 80 edges.\label{fig:botRTbot_net}}
\end{figure}

\begin{figure}
\centering
\includegraphics[width=0.5\textwidth]{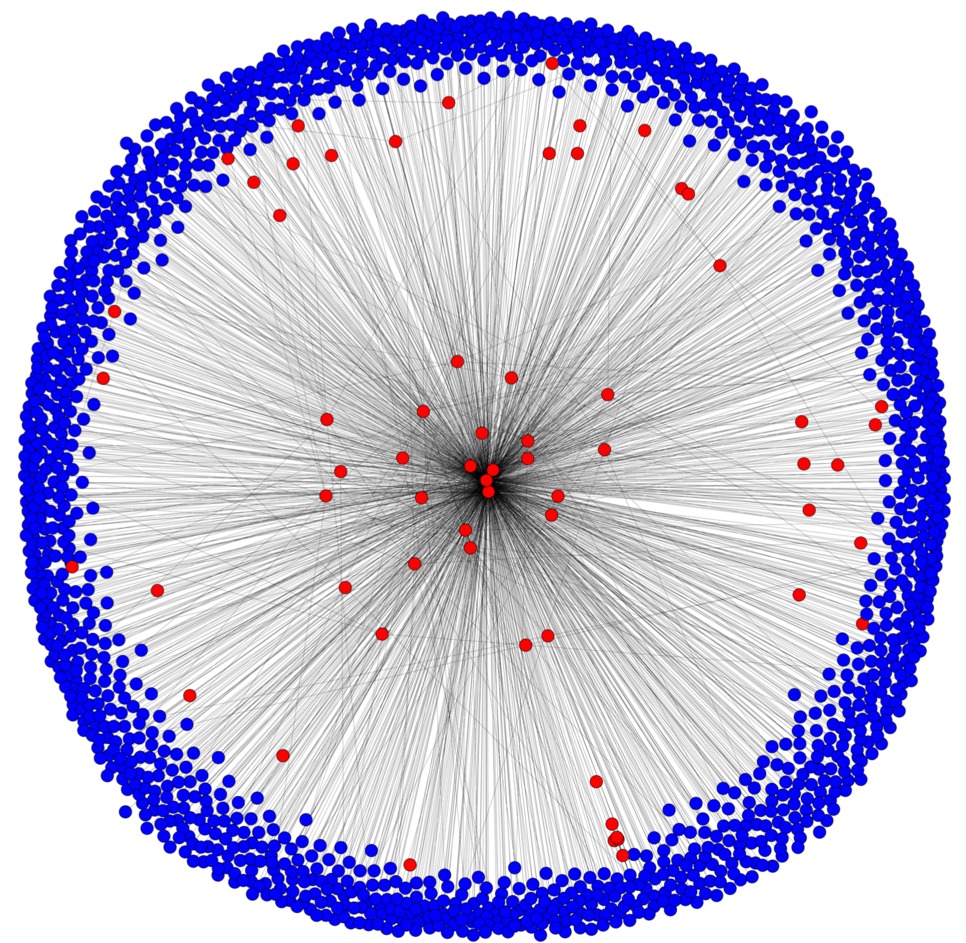}
\caption{Retweet network for \#Tanhuato obtained from our sample, for only humans retweeting bots. All edges are shown, all of the nodes in the outer circle have no connecting edge. This network is composed of 1550 nodes, 1596 edges.\label{fig:humRTbot_net}}
\end{figure}

In figure \ref{fig:RT_net} we show the entire retweet network for our collection. It can be seen that very few bot accounts are responsible for a large proportion of the retweets by humans. This last point is also clear in figure \ref{fig:humRTbot_net}, where only the retweets of bot tweets by humans are shown. Here the central nodes with high valency are the accounts that were retweeted most by humans. In contrast figure \ref{fig:botRTbot_net} shows that bots did not actually retweet themselves much. In fact most bot accounts lie in the outer circle, edgelessly isolated.

The total number of tweets created by bots were 4153, this number represents the
19.9146\% of all tweets.
In total 12905 of all tweets are retweets. A total of 11895 retweets were done by humans, and
1010 retweets were done by bots.
\\ The number of tweets created by bots and retweeted by humans is: 1450
\\ The number of tweets created by humans and retweeted by bots is: 848
\\ The number of tweets created by bots and retweeted by bots is: 76
\\ The number of tweets created by humans and retweeted by humans is: 9896
\\ There are more humans retweeted (10744) than bots retweeted (1526). There is a difference of
635 retweets:
`all retweets '=humans retweeted + bots retweeted + 635

The `missing' 635 retweets belong to tweets created at previous time (before the fist tweet registe-
red). Fortunately, retweets store the info of the original tweet.
Searching the string {\tt http} in the text of each tweet, we found that 17474 tweets from humans
include web pages, and 4736 tweets from bots include web pages.

% TO HERE NOT IN AIES

\section{Text Analysis}

We extract bag-of-words features represented as TF-IDF (term frequency–inverse document frequency) using \cite{sklearn_api}. We then used Singular Value Decomposition (SVD, also referred to as Latent Semantic Indexing in the context of information retrieval and text mining) to look at the distribution of Tweets on the top singular vectors. While the top singular vectors capture the most variance in the bag-of-words features set, for this corpora the difference between the  bot and human tweets was not clear.  We also redid the analysis by removing Spanish stop words and still did not find any discrimination between bots and humans. 

% NEXT PARRAGRAPH NOT IN AIES

\begin{figure}
\centering
\includegraphics[height=.5\textheight]{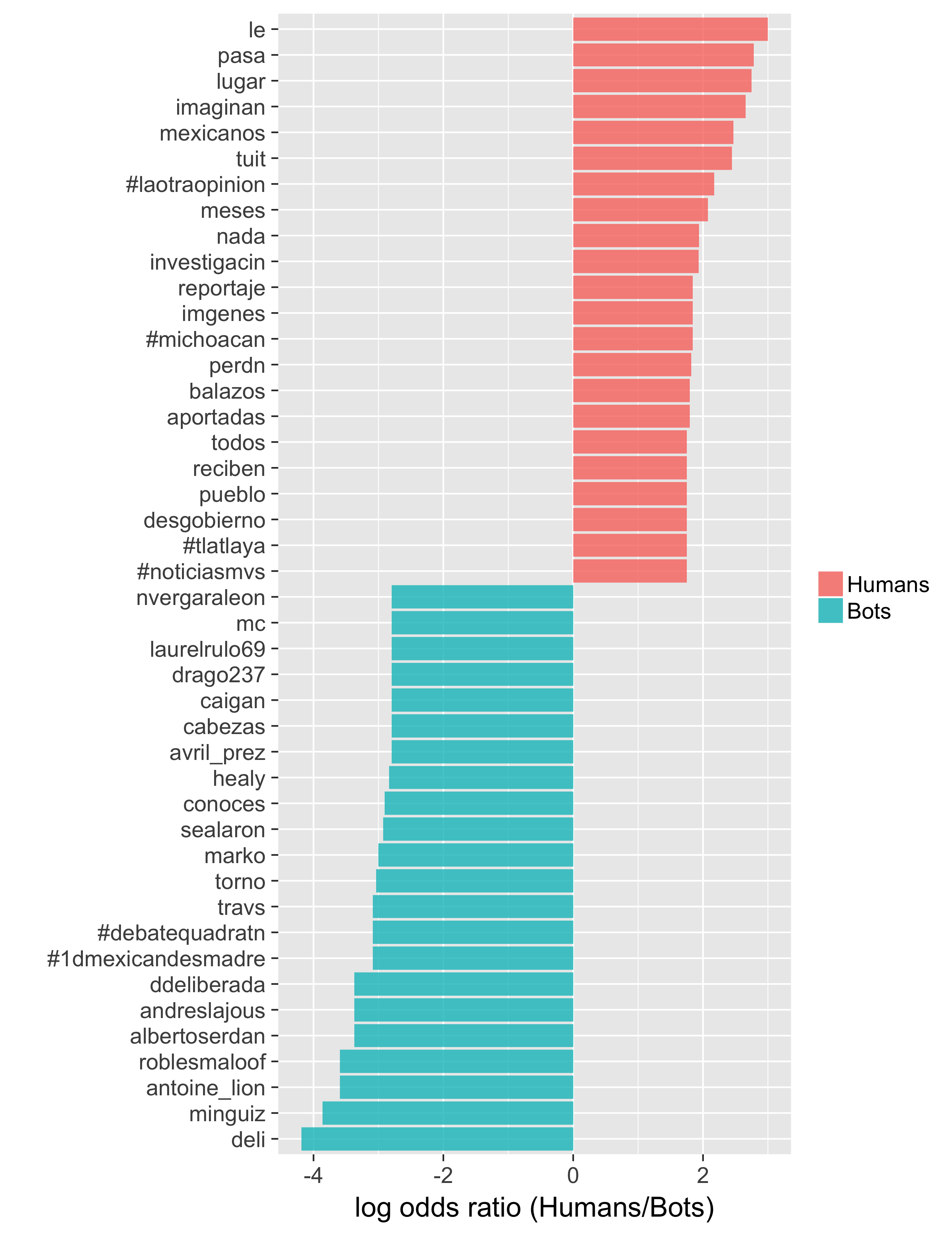}
\caption{Most discriminating words between Bots versus humans as computed by likelihood ratios.\label{fig:bot-likelihoods}} 
\end{figure}

However, as seen in Figure \ref{fig:bot-likelihoods}, by computing the log-odds ratio of the counts of words between the human and bot cohorts (as was done in \cite{tidy_text} for discriminating between two Tweet corpora), we see several terms that are discriminating. Thus, although the bag-of-words features do not capture strong discrimination between bots and humans, the two cohorts are clearly different (specific word usages among bots can be different orders of magnitude since the horizontal axis in Figure \ref{fig:bot-likelihoods} is on a log scale). 

To  better understand the nature of words bots and humans used, we apply  basic sentiment analysis using LabMT \cite{dodds2011temporal}. As discussed in \cite{dodds2011temporal}, the  top 10,000 Spanish  words were presented to Amazon Mechanical Turk where 50  workers rated the  happiness of each word on a scale of 1 to 9 (where 1 is least happy, 9 is most happy, and 5 is neutral). Using these scores for each word, we  compute the average sentiment, $h_{avg}$ for the human and bot corpora using Equation 1 in \cite{dodds2011temporal}. As discussed in  \cite{dodds2011temporal} however,  a great deal of words may have neutral sentiment  (and are essentially commonly used stop words), and the average sentiment score may be biased heavily towards the neutral score of 5.0.  Therefore,  the authors suggest removing words that are within $\Delta h_{avg}$ of 5.0 so that words with stronger sentiment remain.  By selecting an appropriate  $\Delta h_{avg}$, we can remove stop words in a systematic way that does not contribute to sentiment.

\begin{figure}
\centering
\includegraphics[width=\textwidth]{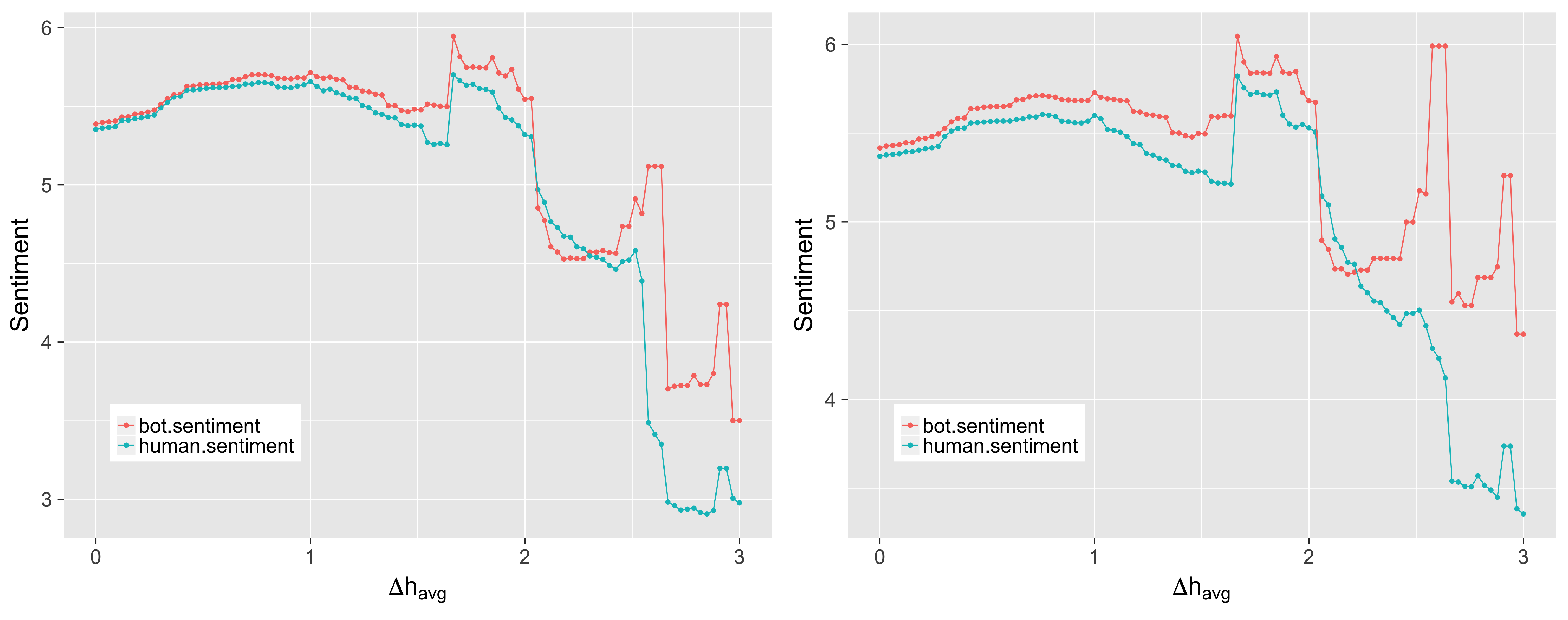}
\caption{\textbf{Left:} Sentiment on Tweets using LabMT. As we filter out neutral words with the $\Delta h_{avg}$, we see that the sentiment from human is significantly lower than bots. \textbf{Right:} Sentiment on Tweets with retweets removed using LabMT. Again, as we filter out neutral words with the $\Delta h_{avg}$, we see that the sentiment from human is significantly lower than bots. However, the correlation between the human and sentiments is much lower when retweets are removed.\label{fig:labmt}} 
\end{figure}

It is not clear what value to select for $\Delta h_{avg}$. While the authors in \cite{dodds2011temporal} suggest $0.5 \leq \Delta h_{avg} \leq 2.5$, here we compute the average sentiment score  $\Delta h_{avg}$ for  $0 \leq \Delta h_{avg} \leq 3.0$ for a more complete understanding. Figure \ref{fig:labmt}, left panel, shows how the tweets average sentiment  changes as we filter out more neutral words. As the neutral words are filtered, we see that the average sentiment is pulled down significantly. This is to be expected as most tweets  are expressing words related to violence.  Interestingly, however, the bots seem to be less emotional than the humans in that their average sentiment is consistently above humans regardless of what $\Delta h_{avg}$ value we use. %This suggests that the  bots are using less emotionally charged language. 

% NEXT FIGURE NOT IN AIES

To investigate this hypothesis further,  we removed all retweets and recomputed the average sentiments. Figure \ref{fig:labmt}, panel on the right, shows again that removing the  retweets does not  change the fact that filtering neutral words yields more negative words.  However, we see that the bot sentiment does not correlate strongly with the human tweets. In other words, as we filter  more neutral words, the human tweets become more negative as before. But the bot tweets remain closer to being neutral. These findings all suggest that the bots were using less emotionally charged words than humans. In other words, it appears that the purpose of the bots in this case was to only distribute information in a  non-sensational manner rather than purposefully stir up emotions.

%Negativity
In addition to using LabMT, we also hand coded a list of negative words, extracted from the corpus of collected tweets, and used it to compare both the bot and human corpora according to the frequency of appearance of words in this list. In order to increase the comparability of these words in a wider volume of tweets, when possible, we suppressed some last letters (that is, we applied ``stemming") such that they could match with different tenses (in case of verbs) and different genders and numbers (in nouns and adjectives) keeping the connotation. We refer to Table \ref{table:list_of_neg_words} for this list of incomplete words.

To check matches between words in Table~\ref{table:list_of_neg_words} and the text in tweets, we remove URLs from the text in tweets, replace non-ASCII characters (like ``\~n '', stressed vowel \'a,\'e,\'i,\'o,\'u and ``?` '') by their ASCII equivalent (``n'',a, e, i, o, u,``?''). We also transform all capital letters to lowercase. The transformed text were split into single words to compare individually. In order to increase comparison speeds, we group the words alphabetically and compare only with words starting with the same letter, skipping also words starting with symbols, numbers. Finally, we only check if the words in Table~\ref{table:list_of_neg_words} with the same initial letter as each word in split message starts with the same letters.

To prevent a misplaced punctuation mark from not matching a word, a second analysis was performed suppressing the first letter in each word, and checking if this shorter word matches with Table~\ref{table:list_of_neg_words}. This analysis also reveals no difference. Our method of comparison fails when a negative sentiment word is misspelled, but one expects that the sentiment of the tweet remains congruent in the whole text. Then, if the text is long, we are more likely to find another negative word but spelled correctly. Conversely, short texts are more likely to have less misspelled words.

\begin{table}[htbp]
\setlength{\tabcolsep}{8pt}
\begin{center}
\begin{tabular}{llll}
arma & 	 culpable & 	 jodid$\ast$ & 	 sanguinari$\ast$  \\
 asesin$\ast$ & 	 delincuen$\ast$ & 	 levanton & 	 secuestro  \\
 asesinat$\ast$ & 	 dispara & 	 maltrat$\ast$ & 	 tortura \\
 bala & 	 disparos & 	 masacre & 	 violacion \\
 balazo & 	 ejecucion & 	 matanza & 	 violenta\\
 brutal & 	 ejecut$\ast$ & 	 matar & 	\\
 cartel & 	 exterminio & 	 mentir & 	\\
 castigo & 	 fals$\ast$ & 	 muerte & 	\\
 corrupcion & 	 genocidio & 	 pistola & 	\\
 corrupt & 	 guerra & 	 represion & 	\\
 crimen & 	 incendia & 	 represiv$\ast$ & 	\\
 criminal & 	 jode$\ast$ & 	 sangriento & 	\\
\end{tabular}
\caption{List of negative feeling words (an $\ast$ is placed when letters can omitted without changes in connotation.}
\label{table:list_of_neg_words}
\end{center}
\end{table}

\begin{figure}
\centering
\includegraphics[width=.75\textwidth]{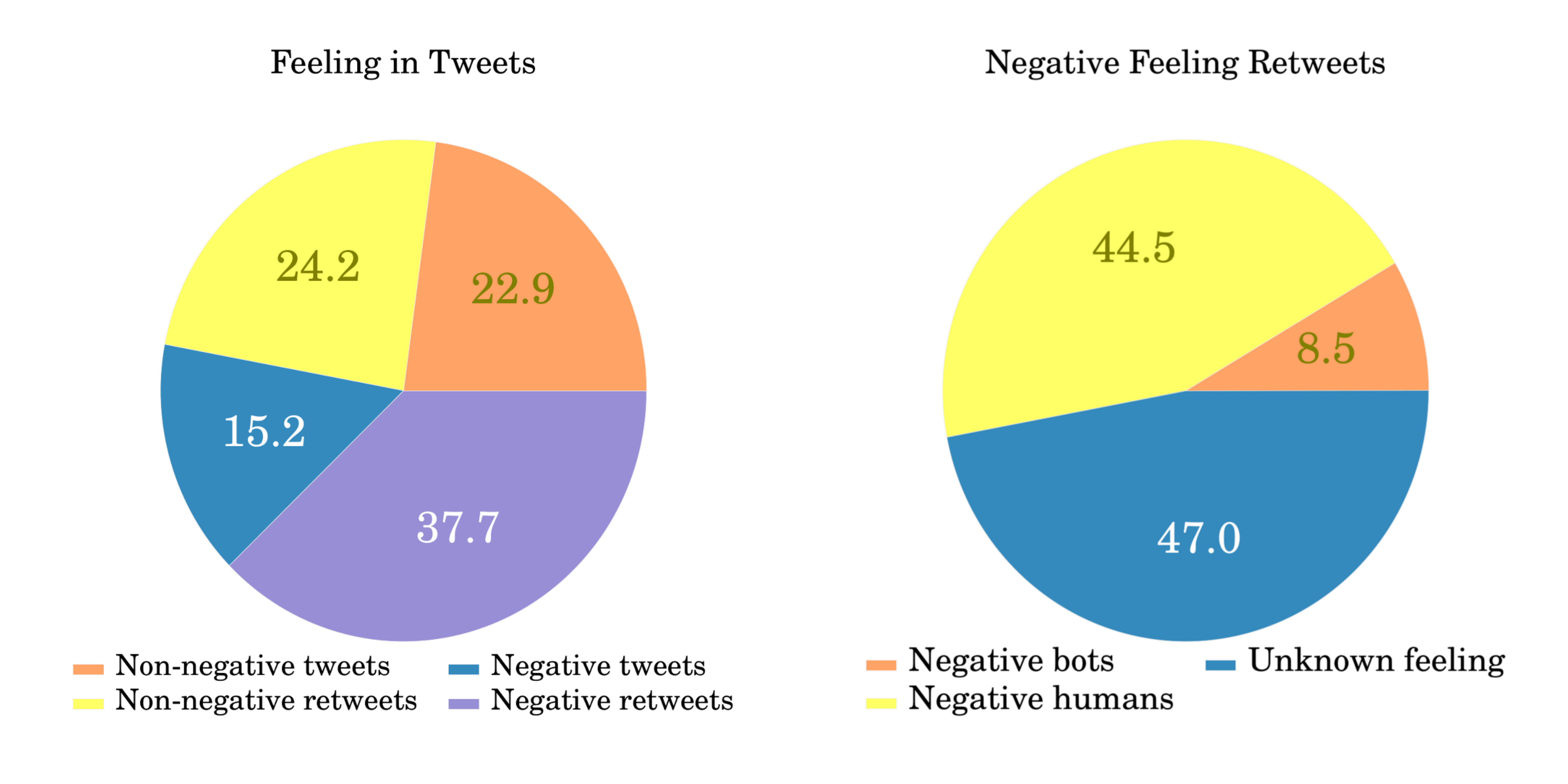}
\caption{The total volume of twitter texts were comparing with words in Table~\ref{table:list_of_neg_words}. {\sc left: } Tweet classification in {\bfseries Negative} and {\bfseries Non-negative}. {\sc right} Percentage of negative feeling texts by user type.\label{fig:pie2}}
\end{figure}

To distinguish what kind of information is most shared, we consider the total of tweets and assign a numerical value to each one. This value was initialized in 0 increased by a constant, depending on the number of matches with the~Table~\ref{table:list_of_neg_words}. Assuming that a tweet has a negative feeling when its value is different to zero, we show in Figure~\ref{fig:pie2} that the largest volume of tweets comes from retweets with a negative feeling text. A closer reading of the entire tweet corpus revealed that the most of the messages which are non-negative cannot be identified as positive or neutral. Their texts share URLs and/or the sentiment cannot be determined by word inspection.

% NEXT FIGURE NOT IN AIES

% \begin{figure}
% \label{fig:negative}
% \centering
% \includegraphics[width=0.5\textwidth]{Negative_feelling.png}
% \caption{Feeling distribution of all twiter texts according to their matching with words in Table~\ref{table:list_of_neg_words}.}
% \end{figure}

\section{Conclusions}
In this work we presented a case study of socialbots for a specific trending topic in Mexican Twitter. While numerous studies have suggested that socialbots act as disrupting agents of information, in our case study we found the opposite. The socialbots were in fact enabling the flow of information to ensure that the report about these atrocities reached the public and information was not stifled. Of course, from the point of the police authorities the socialbots may be viewed as agents of disruption and it is therefore a matter of perspective if socialbot are enablers or not. Our case study suggests that the role and landscape of socialbots is far more complex than simple binary categorizations. Our work highlights the need for further research to understand the ethical implications of such automated social activity.

\section*{Acknowledgments.} We thank IPAM in UCLA and the organizers of the Cultural Analytics program, CNetS and the BotOrNot team in IU, and also Twitter for allowing access to data through their APIs. PSS acknowledges support from UNAM-DGAPA-PAPIIT-IN102716 and UC-MEXUS CN-16-43.

\bibliographystyle{plain}% siam 
%\nocite{*}
%\bibliography{biblio2.bib}
%\bibliography{biblio.bib}

\end{document}